\documentstyle{article}

\def\abstract#1{\vskip 7mm 
        \begin{center}{\large Abstract}\par \smallskip
                \begin{minipage}[c]{12cm}
                        \small #1
                \end{minipage}
        \end{center}
}
\def\title#1{\begin{center}{\Large\bf #1}\end{center}}
\def\author#1{\vskip 5mm \begin{center}{#1}\end{center}}
\def\address#1{\begin{center}{\it #1}\end{center}}

\newtheorem{theorem}     {Theorem}
\newtheorem{lemma}       {Lemma}

\newtheorem{definition}  {Definition}

\newtheorem{remark}      {Remark}
 
\begin{document}
\title{Global properties of 
higher-dimensional 
cosmological spacetimes}

\author{Makoto Narita} 

\address{Max-Planck-Institut f\"ur Gravitationsphysik, \\
Albert-Einstein-Institut, 
Am M\"uhlenberg 1, 
D-14476 Golm, \\
Germany\\
E-mail: maknar@aei-potsdam.mpg.de}

\abstract{We study global existence problems and asymptotic behavior of 
higher-dimensional inhomogeneous spacetimes with a compact Cauchy surface 
in the Einstein-Maxwell-dilaton (EMD) system. 
Spacelike $T^{D-2}$-symmetry is assumed, where $D\geq 4$ is 
spacetime dimension. The system of the evolution equations of the 
EMD equations 
in the areal time coordinate is reduced to a wave map system, 
and a global existence theorem for the system is shown. As a corollary of 
this theorem, a global existence theorem in the constant mean curvature time 
coordinate is obtained. Finally, 
for vacuum Einstein gravity in arbitrary dimension, 
we show existence theorems of asymptotically 
velocity-terms dominated singularities 
in the both cases which free functions are analytic and smooth.}

\section{Introduction and summary}
\label{intro}
Global existence problems of fundamental field equations must 
be solved as the first step of 
the strong cosmic censorship, which states that 
generic Cauchy data sets have maximal Cauchy developments 
which are locally inextendible as Lorentzian manifolds. 
It is important to consider the strong cosmic censorship 
in scope of unified theories such that 
superstring/M-theory. 
As is well known, such unified theories predict that the spacetime 
dimension exceeds four and it is expected that such 
extra dimensions must not be different from ordinary dimensions 
and should be taken into account in the asymptotic regions 
(e.g. near the singularities) of spacetimes. 
Although higher-dimensional model have a long history 
within superstring/M-theory, mathematically rigorous results for 
cosmological spacetimes are much less understood 
than for stationary ones (for examples, see~\cite{GSW}). 
In particular, 
there has been less exploration of global {\it in time} problems 
in higher-dimensional, cosmological spacetimes.  
Recently, spatially homogeneous cosmological models was analyzed~\cite{HS}. 
Then, we want to study an inhomogeneous cosmological case as the next step.

In the cosmological context, 
it seems convenient to consider spacetimes which develop 
from smooth Cauchy data given on a compact, connected and orientable 
Cauchy hypersurface. 
In addition, we would like to investigate inhomogeneous cosmological 
spacetimes with dynamical 
degree of freedom of gravity. 
Then, we will consider globally hyperbolic spacetimes 
$({\cal M}_{D},g)$, with ${\cal M}_{D}={\cal M}_{D-1}\times {\rm R}$ 
a smooth $D$-dimensional manifold ($D\geq 4$), $g$ a Lorentzian metric,  
which develop from smooth Cauchy data 
invariant under an effective action of 
$G_{D-2}=U(1)\times U(1)\times\cdots\times U(1)=T^{D-2}$ 
on a compact $(D-1)$-dimensional spacelike manifold ${\cal M}_{D-1}$. 
The same symmetry is assumed for matter fields if they exist. 
The resulting system of field equations becomes one of 
1+1 nonlinear partial differential equations (PDEs). 
From Mostert's theorem~\cite{MP} it is admitted that ${\cal M}_{D-1}/G_{D-2}$ 
is a circle and 
${\cal M}_{D-1}$ is homeomorphic to $S^{1}\times T^{D-2}= T^{D-1}$. 
Therefore, 
we will suppose ${\cal M}_{D-1}\approx T^{D-1}$.

In four-dimensional vacuum spacetimes with the above setting, 
Moncrief and Isenberg have proved global existence theorems of 
the Einstein equations 
in areal and constant mean curvature time coordinates~\cite{MV,IM}. 
Recently, it has been generalized to 
non-vacuum case~\cite{AH,ARW,HO,NM02,NM03}.
One purpose of the present paper is to extend the previous results 
to higher-dimensional spacetimes.

Once it has been shown global existence theorems, 
we want to know asymptotic behavior of cosmological spacetimes as the next 
step. 
That is, it should be analyzed nature of spacetime regions near singularity 
(if incomplete) and 
near infinity (if complete). 
In this paper, we will focus to consider nature of singularities. 
It is conjectured that, 
near a cosmological singularity, a decoupling of spatial points occurs 
in the sense that evolution equations 
cease to be PDEs to simply become, 
at each spatial point asymptotically, ordinary differential 
equations with respect to time~\cite{BKL}. 
In other words, cosmological singularities are locally 
asymptotically velocity-terms dominated (AVTD
\footnote{In this article, the word ``AVTD behavior'' is equivalent to 
``Kasner-like behavior''.}) or 
mixmaster. Concerning this BKL conjecture, 
it has been shown that four-dimensional Gowdy symmetric spacetimes with or 
without stringy matter fields (with dilaton couplings)
have AVTD singularities in general in the sense 
that the singular solutions depend on the maximal number 
of arbitrary functions~\cite{KR,NTM}.
These results have been made no-symmetric and higher-dimensional 
generalization~\cite{DHRW}. 
That paper has established AVTD behavior for vacuum gravity 
in spacetime dimension $D>10$ and 
for the Einstein-dilaton-matter system with dilaton couplings 
in spacetime dimension $D\geq 2$. 
Thus, the question whether the AVTD behavior exists or not 
still remains open for vacuum Einstein gravity 
in dimensions $D\in[5,10]$ rigorously.
Another purpose of the present paper is to show that 
there is AVTD behavior for vacuum Einstein gravity 
in arbitrary spacetime dimension. 
This result complements the BKL conjecture by combining together the previous 
works mentioned above.

The above results concerning AVTD behavior are of $C^{\omega}$ (analytic) 
category. Recently, it has been extended to $C^{\infty}$ (smooth) category 
in the case of four-dimensional vacuum Gowdy spacetimes with any spatial 
topology~\cite{RA,SF}. 
We will generalize this to higher-dimensional spacetimes and tie the result 
on AVTD singularities together with the global existence results.

Let us consider a truncated action of the bosonic supergravity theory 
(i.e. low energy effective superstring/M-theory), 
which contains only 
the gravitational (metric), the dilaton and $p$-form fields. 
It can be shown that this truncation is consistent, 
in the sense that the fields that are retained are not sources 
for the fields that are eliminated~\cite{SK}. 
In this paper, we will consider the case $p=1$. 
Thus, the system becomes the Einstein-Maxwell-dilaton (EMD) system. 
The action for the $D$-dimensional EMD system is given by 
\begin{eqnarray}
\label{d-action}
S_{D}=\int d^{D}x\sqrt{-{\rm det}g}\left[-{}^DR+2(\partial\phi)^2
+e^{-2a\phi}F^2\right],
\end{eqnarray}
where ${}^DR$ is the Ricci scalar with respect to $g$, 
$\phi$ is the dilaton field, 
$F_{\mu\nu}=\partial_{\mu}A_{\nu}-\partial_{\nu}A_{\mu}$ is 
the Maxwell field strength 
and $a$ is a coupling constant. 
Varying the action we have the following field equations:
\begin{eqnarray}
\label{einstein-eq}
{}^DR_{\mu\nu}=2\partial_{\mu}\phi\partial_{\nu}\phi
+e^{-2a\phi}\left[2g^{\lambda\sigma}F_{\mu\lambda}F_{\nu\sigma}
-\frac{1}{D-2}g_{\mu\nu}F^2\right],
\end{eqnarray}
\begin{eqnarray}
\label{dilaton-eq}
{}^D\Box\phi+\frac{a}{2}e^{-2a\phi}F^2=0,
\end{eqnarray}
\begin{eqnarray}
\label{maxwell-eq}
\partial_{\mu}(\sqrt{-{\rm det}g}e^{-2a\phi}F^{\mu\nu})=0
\end{eqnarray}
\begin{eqnarray}
\label{bianchi-id}
\partial_{[\mu}F_{\nu\lambda]}=0,
\end{eqnarray}
where $\mu$, $\nu$, $\lambda$ run from $0$ to $D-1$, 
${}^DR_{\mu\nu}$ and ${}^D\Box$ are the Ricci tensor 
and the d'Alembertian 
with respect to $g_{\mu\nu}$, respectively.
Note that truncation of the Maxwell fields or of 
both the Maxwell and the dilaton fields is consistent, 
but trivializing only the dilaton field does not induce the Einstein-Maxwell 
(EM) system from the the EMD system. In this case, 
as we can see from the dilaton equation~(\ref{dilaton-eq}), 
the EM system with a constraint $F^2=0$ would be obtained.

\section{$G_{D-3}$-invariant spacetimes: 
Reduction of a wave map system coupled to three-dimensional gravity}
We take ${\cal M}_{D-1}$ to be a principal fiber bundle with compact, 
two-dimensional base $\Sigma$, 
with fibers the orbits of $G_{D-3}=U(1)\times\cdots\times U(1)=T^{D-3}$, 
and with a metric $g$ on ${\cal M}_{D}$ invariant 
under the right action of $G_{D-3}$. 
Therefore, ${\cal M}_{D}$ is a principal fiber bundle with base 
$\Sigma\times {\rm R}$ and group $G_{D-3}$. 

Let $\xi_I=\partial_I$ $(I=3,\cdots,D-1)$ be $(D-3)$ spacelike
commuting Killing vector fields,
such that ${\cal L}_{\xi_I}g=0$ and $[\xi_I,\xi_J]=0$. 
Note that the Killing vectors are tangent to ${\cal M}_{D-1}$, 
hence spacelike. 
Form the hypotheses, the metric $g$ for the spacetime is~\cite{CBDWM}
\begin{eqnarray}
g=f^{-1}\gamma_{mn}dx^mdx^n+f_{IJ}\Theta^{I}\Theta^{J},\hspace{.5cm}
\Theta^{I}:=dx^I+W^I{}_mdx^m,
\end{eqnarray}
where $f^{-1}\gamma_{mn}$ is a metric on $\Sigma\times {\rm R}$, 
$m,n,i,j=0,1,2$, $f={\rm det}f_{IJ}>0$. 
$\gamma_{mn}$, $W^I{}_m$ and $f_{IJ}$ are depend only on the 
coordinates $x^{m}$, that is a necessary and sufficient condition 
for the metric $g$ to be 
invariant under the right action of $G_{D-3}$.

We consider the Maxwell field $A_{\mu}$.
Assume that there is a gauge such that the $U(1)$ gauge field respects 
the spacetime symmetry
${\cal L}_{\xi_I}A=0$. 
Then we have
$\partial_{I}A_{\mu}=0$.
According to \cite{IW} (see also~\cite{IU}), we can define the {\it Maxwell field potentials}  
$\Phi_{I}$ and $\Psi$ 
from equations (\ref{bianchi-id}) and (\ref{maxwell-eq}) as follows.
\begin{eqnarray}
F_{\mu I}&=&\partial_{\mu}\Phi_I,
\label{def-Phi}\\
e^{-2a\phi}F^{mn}&=&\frac{2f}{\sqrt{-{\rm det}\gamma}}\epsilon^{mn\mu}\partial_{\mu}\Psi,
\label{def-Psi}
\end{eqnarray}
where $\epsilon_{mn\mu}$ denotes 
the permutation symbol such that $\epsilon_{012}=1$ and 
$\epsilon_{mnI}=0$.
The equation (\ref{def-Phi}) satisfies a part 
of the Bianchi identity (\ref{bianchi-id}), 
while the 
equation (\ref{def-Psi}) fulfills a part 
of the Maxwell equations (\ref{maxwell-eq}). 
From (\ref{def-Phi}) and (\ref{def-Psi}) the remaining Maxwell equations 
and the Bianchi identity 
can be described through 
$\Phi_I$, $\Psi$, $f_{IJ}$, $W^{I}_{m}$ and $\gamma_{mn}$.  
Introduce {\it torsion} defined by
\begin{eqnarray}
\omega_{I\mu}&=&\sqrt{-{\rm det}g}\epsilon_{\mu\mu_3\cdots\mu_{D-1}\nu\lambda}
\xi_3^{\mu_3}\cdots\xi_{D-1}^{\mu_{D-1}}\partial^\nu\xi_I{}^\lambda\nonumber \\
&=&ff_{IJ}\sqrt{-{\rm det}\gamma}\epsilon_{ij\mu}\gamma^{im}
\gamma^{jn}\partial_mW^J{}_n
\label{def-omega}
\end{eqnarray}
where $\epsilon_{\mu_0\cdots\mu_{D-1}}$ is the Levi-Civita tensor $\epsilon_{0\cdots D-1}=1$ 
and $\gamma^{mn}$ is the inverse metric of $\gamma_{mn}$.
One can obtain the following evolution equations 
for the Maxwell field potentials $\Phi_{I}$ and $\Psi$ 
from equations (\ref{maxwell-eq}) and (\ref{bianchi-id}):
\begin{eqnarray}
\Box
\Phi_I=f^{JK}\nabla^{m} f_{IJ}\nabla_{m}\Phi_K
-f^{-1}\nabla^{m}\Psi\omega_{Im}e^{2a\phi}+2a\nabla^{m}\phi\nabla_{m}\Phi_{I},
\label{Phi-evolution-eq'}
\end{eqnarray}
and
\begin{eqnarray}
\Box\Psi=f^{-1}\nabla^{m} f\nabla_{m}\Psi
+f^{IJ}\nabla^{m}\Phi_I\omega_{Jm}e^{-2a\phi}-2a\nabla^{m}\phi\nabla_{m}\Psi,
\label{Psi-evolution-eq'}
\end{eqnarray}
where $\nabla$ and $\Box=\nabla^{m}\nabla_{m}$ are 
the covariant differential operator 
and the d'Alembertian of $\gamma_{mn}$, respectively. 

The equations for the dilaton field takes the form
\begin{eqnarray}
\label{phi-evolution-eq}
\Box\phi=-\frac{a}{2}(e^{-2a\phi}f^{IJ}\nabla^{m}\Phi_{I}\nabla_{m}\Phi_{J}
-e^{2a\phi}f^{-1}\nabla^{m}\Psi\nabla_{m}\Psi).
\end{eqnarray}

Now, we will write down the  Einstein equations. 
From the mixed $(I,m)$-components of the Einstein 
equations (\ref{einstein-eq}) we have 
\begin{eqnarray}
\label{mixed}
\partial_{\mu}\omega_{I\nu}-\partial_{\nu}\omega_{I\mu}
=2(\partial_{\nu}\Phi_{I}\partial_{\mu}\Psi-\partial_{\mu}\Phi_{I}\partial_{\nu}\Psi). 
\end{eqnarray}
From this, we can define the {\it twist potential} $Q_{I}$ such that 
\begin{equation}
\label{def-q}
dQ_{I}+H_{I}=\omega_{I}+\Phi_{I}d\Psi-\Psi d\Phi_{I},
\end{equation}
where $H_{I}$ is a harmonic one-form for some given Riemannian metric 
on $\Sigma$ which is compact~\cite{CJ,CBM96}. 
For simplicity, we will suppose 
\begin{eqnarray}
\label{harmonic-1-form}
H_{I}\equiv 0. 
\end{eqnarray}
Evolution equations for $Q_{I}$ can be found by taking a three-covariant divergence 
of equation (\ref{def-q}) taking account definition of $\omega_{I\mu}$ (\ref{def-omega})
\begin{eqnarray}
\Box Q_I&=&
f^{-1}(\nabla^{m} f+\Psi \nabla^{m}\Psi)
(\nabla_{m} Q_I+\Psi \nabla_{m}\Phi_I-\Phi_I \nabla_{m}\Psi)
\nonumber\\
&&{}
+f^{JK}(\nabla^{m} f_{IJ}+\Phi_I \nabla^{m}\Phi_J)
(\nabla_{m} Q_K+\Psi \nabla_{m}\Phi_K-\Phi_K \nabla_{m}\Psi)
\nonumber\\
&&{}
+f^{-1}\Phi_I \nabla^{m} f\nabla_{m}\Psi-f^{JK}\Psi \nabla^{m} f_{IJ}\nabla_{m}\Phi_K.
\label{q-evolution-eq}
\end{eqnarray}

Next, evolution equations for the {\it scalar multiplet} $f_{IJ}$ follow from 
$(I,J)$-components of the Einstein 
equations (\ref{einstein-eq}):  
\begin{eqnarray}
\Box f_{IJ}&=&f^{KL}\nabla^{m} f_{IK}\nabla_{m} f_{JL}
-f^{-1}(\nabla^{m} Q_I+\Psi \nabla^{m}\Phi_I-\Phi_I \nabla^{m}\Psi)
(\nabla_{m} Q_J+\Psi \nabla_{m}\Phi_J-\Phi_J \nabla_{m}\Psi)
\nonumber\\
&&{}
-2e^{-2a\phi}\nabla^{m}\Phi_I\nabla_{m}\Phi_J
+\frac{2}{D-2}f_{IJ}(e^{-2a\phi}f^{KL}\nabla^{m}\Phi_K\nabla_{m}\Phi_L-e^{2a\phi}f^{-1}\nabla^{m}\Psi\nabla_{m}\Psi).
\label{f-evolution-eq}
\end{eqnarray}

The effective Einstein equations on the three-spacetime whose metric is $\gamma$ follow from $(m,n)$ and $(I,J)$-components 
of the Einstein 
equations (\ref{einstein-eq}):
\begin{eqnarray}
{}^{\gamma}R_{mn}&&=\frac{1}{4}(f^{-2}\nabla_{m}f\nabla_{n}f
+f^{IJ}f^{KL}\nabla_{m}f_{IK}\nabla_{n}f_{JL})
+e^{-2a\phi}f^{IJ}\nabla_{m}\Phi_{I}\nabla_{n}\Phi_{J}
+e^{2a\phi}f^{-1}\nabla_{m}\Psi\nabla_{n}\Psi
\nonumber\\
&&{}+\frac{1}{2}f^{-1}f^{IJ}
(\nabla_{m}Q_{I}+\Psi \nabla_{m}\Phi_{I}-\Phi_I \nabla_{m}\Psi)
(\nabla_{n}Q_{J}+\Psi \nabla_{n}\Phi_{J}-\Phi_J \nabla_{n}\Psi)
+2\nabla_{m}\phi\nabla_{n}\phi,
\label{3einstein-eq}
\end{eqnarray}
where ${}^{\gamma}R_{mn}$ is the Ricci tensor for $\gamma_{mn}$.

Note that the Maxwell equations (\ref{Phi-evolution-eq'}) and (\ref{Psi-evolution-eq'}) can be rewritten by using $Q_{I}$ as follows:
\begin{eqnarray}
\Box\Phi_I=f^{JK}\nabla^{m} f_{IJ}\nabla_{m}\Phi_K
-f^{-1}\nabla^{m}\Psi(\nabla_{m} Q_I+\Psi \nabla_{m}\Phi_I-\Phi_I \nabla_{m}\Psi)e^{2a\phi}+2a\nabla^{m}\phi\nabla_{m}\Phi_{I},
\label{Phi-evolution-eq}
\end{eqnarray}
and
\begin{eqnarray}
\Box\Psi=f^{-1}\nabla^{m} f\nabla_{m}\Psi
+f^{IJ}\nabla^{m}\Phi_I(\nabla_{m} Q_J+\Psi \nabla_{m}\Phi_J-\Phi_J \nabla_{m}\Psi)e^{-2a\phi}-2a\nabla^{m}\phi\nabla_{m}\Psi.
\label{Psi-evolution-eq}
\end{eqnarray}

When $f_{IJ}$, $\gamma_{mn}$, $Q_{I}$, $\Phi_{I}$, $\Psi$ and $\phi$ are known on $\Sigma\times{\rm R}$, 
we can get two-forms ${\cal W}^{I}$ on $\Sigma\times{\rm R}$, where
\begin{equation}
{\cal W}_{ij}^{I}=\partial_{i}W^I{}_{j}-\partial_{j}W^I{}_{i}=-f^{-1}f^{IJ}\sqrt{-{\rm det}\gamma}\epsilon_{ijm}
\gamma^{mn}(\partial_{n}Q_{J}+\Psi \partial_{n}\Phi_{J}-\Phi_J \partial_{n}\Psi).
\end{equation}
We can deduce from ${\cal W}^{I}$ one-forms $\Theta^{I}$ on ${\cal M}_{D}$ if and only if  
the following two conditions hold~\cite{CJ,CBM96}. 
One is a local condition which is the system of evolution equations. 
Another is a global condition. 
Since each $U(1)$-symmetric one-form $\Theta^{I}$ is independent, 
in the case $\Sigma$ compact, $G_{D-3}=T^{D-3}$, the global condition reads 
\begin{equation}
\label{integral}
n_{I}=\frac{1}{2\pi}\int_{\Sigma}d\Theta^{I} =\frac{1}{2\pi}\int_{\Sigma}{\cal W}^{I},\hspace{.5cm}(n_{I}\in {\rm Z}).
\end{equation}

In summary, 
the following action describing a wave map 
coupled to three-dimensional gravity is obtained as an effective action:  
\begin{eqnarray}
\label{action}
S_{3}&=&\int d^3x \sqrt{-{\rm det}\gamma}\left[
-{}^{\gamma}R
+\frac{1}{4}f^{-2}\nabla^{m}f\nabla_{m}f
+\frac{1}{4}f^{IJ}f^{KL}\nabla^{m}f_{IK}\nabla_{m}f_{JL}
\right.
\nonumber\\
&&{}
\left.
+e^{-2a\phi}f^{IJ}\nabla^{m}\Phi_I\nabla_{m}\Phi_J
+e^{2a\phi}f^{-1}\nabla^{m}\Psi\nabla_{m}\Psi+2\nabla^{m}\phi\nabla_{m}\phi
\right.
\nonumber\\
&&{}
\left.
+\frac{1}{2}f^{-1}f^{IJ}
(\nabla^{m}Q_I+\Psi\nabla^{m}\Phi_I-\Phi_I \nabla^{m}\Psi)
(\nabla_{m}Q_J+\Psi\nabla_{m}\Phi_J-\Phi_J \nabla_{m}\Psi)
\right],
\end{eqnarray}
where ${}^{\gamma}R$ is the Ricci scalar for $\gamma_{mn}$. 
It is easy to show that the evolution equations, (\ref{phi-evolution-eq}), 
(\ref{q-evolution-eq}), (\ref{f-evolution-eq}), 
(\ref{Phi-evolution-eq}), (\ref{Psi-evolution-eq}), 
and the three-dimensional Einstein equations (\ref{3einstein-eq}) are obtain 
from this action~(\ref{action}).
The only wave map consisting of scalar fields has dynamical degrees of freedom, 
but three-dimensional gravity does not have ones.

\section{Global existence theorem for $T^{D-2}$-symmetric spacetimes 
in areal time coordinate}
We will consider $G_{D-2}=U(1)\times\cdots\times U(1)=T^{D-2}$-invariant cosmological spacetimes. That is, 
we assume the existence of another spacelike Killing vector field $\xi_2=\partial_2$ 
which commutes with
the other Killing vectors $[\xi_2,\xi_I]=0$.
The Maxwell and the dilaton fields are also assumed to be independent of the coordinate $x^2$. 
Thus, ${\cal M}_{D-1}$ can be parametrized by $x^{{\cal I}}$ and 
$x^{1}=\theta\in{\cal M}_{D-1}/G_{D-2}\approx [0,2\pi]\vert_{{\rm mod}2\pi}\approx S^{1}$, 
where ${\cal I}=2,\cdots,D-1$. 
The Frobenius integrability condition is~\cite{DFC}
\begin{eqnarray}
\label{frobenius}
\xi_2{}^{[\mu_2}\cdots\xi_{D-1}{}^{\mu_{D-1}}\partial^\nu\xi_{\cal I}{}^{\lambda]}
=0.
\end{eqnarray}
Spacetimes satisfying the above condition admit a foliation by 
two-dimensional integrable surfaces orthogonal to 
the Killing fields $\xi_{{\cal I}}$. 
From equation (\ref{def-omega}) one can obtain that the condition (\ref{frobenius}) for ${\cal I}=I$ is 
\begin{equation}
\omega_{I2}=0.
\label{omega2}
\end{equation}
The equations (\ref{omega2}) are satisfied if they hold somewhere 
at one point in the spacetimes. 
Indeed, we have the following equation by (\ref{mixed}):
\begin{equation}
\partial_{\mu}\omega_{I2}=2(\partial_{2}\Phi_{I}\partial_{\mu}\Psi-\partial_{\mu}\Phi_{I}\partial_{2}\Psi)=0,
\end{equation}
because of $\partial_{2}\Phi_{I}=\partial_{2}\Psi =0$. 
Since it need not to distinguish $\xi_{2}$ from $\xi_{I}$, the Frobenius integrability condition (\ref{frobenius}) 
is satisfied for ${\cal I}$ if equations (\ref{omega2}) hold at least one point
\footnote{In the case of four-dimensional $U(1)\times U(1)$-symmetric spacetimes, 
it is known that the above 
assumption corresponds to vanishing twist constants in the case of $T^{3}$-spatial topology. 
In addition, if there is a symmetry axis in the spatial section, like $S^{2}\times S^{1}$, 
the condition (\ref{omega2}) holds at the axis, 
then the hypersurface-orthogonality is automatically satisfied~\cite{CP}.}.
Hereafter, we assume this hypersurface-orthogonality condition.

Under the above assumption, 
the line element of the spacetimes takes the $2\times 2$-$(D-2)\times (D-2)$ 
block-diagonal form. 
Thus, the three-dimensional Lorentzian 
metric can be written in the form
\begin{eqnarray}
\gamma_{ij}=e^{2\lambda}(-dt^2+d\theta^2)+\rho^2 d\psi^2,\hspace{.5cm}W^{I}_{0}=W^{I}_{1}=0,
\end{eqnarray}
where $t=x^0$, $\psi=x^2$, $\lambda=\lambda(t,\theta)$ 
and $\rho=\rho(t,\theta)$. 
Here, we have used the fact that two-dimensional spacetime 
(which is spanned by $t$ and $\theta$-coordinates) 
can be conformal flat. 
In this coordinate, we have a linear wave equation for $\rho$ from 
the Einstein equations (\ref{3einstein-eq}) 
for three-spacetime, 
\begin{eqnarray}
(\partial_{t}^2-\partial_{\theta}^2)\rho=0.
\label{rho-evolution-eq}
\end{eqnarray}
By using equation (\ref{rho-evolution-eq}) and 
the same argument with Gowdy~\cite{GR}, 
we can take {\it areal time coordinate} $\rho=t$. 
Then, the spacetime metric is 
\begin{eqnarray}
g=f^{-1}[e^{2\lambda}(-dt^2+d\theta^2)+t^2d\psi^2]
+f_{IJ}(dx^I+W^I d\psi)(dx^J+W^Jd\psi),
\label{areal-time-coordinate}
\end{eqnarray}
where the metric functions depend only on $t$ and $\theta$. 
The same dependence is assumed for functions of matter fields. 
Note that a metric of $T^{3}$-Gowdy symmetric spacetimes would be 
induced from the metric (\ref{areal-time-coordinate}) if $D=4$~\cite{GR}. 

In the areal time coordinate (\ref{areal-time-coordinate}), 
we have constraint equations from (\ref{3einstein-eq}) as follows:
\begin{eqnarray}
2t^{-1}\partial_{t}\lambda
&=&\frac{1}{4}f^{-2}[(\partial_{t}f)^2+(\partial_{\theta}f)^2]
+\frac{1}{4}f^{IJ}f^{MN}(\partial_{t}f_{IM}\partial_{t}f_{JN}+\partial_{\theta}f_{IM}\partial_{\theta}f_{JN})\nonumber \\
&&{}+e^{-2a\phi}f^{IJ}(\partial_{t}\Phi_{I}\partial_{t}\Phi_{J}+\partial_{\theta}\Phi_{I}\partial_{\theta}\Phi_{J})
+e^{2a\phi}f^{-1}[(\partial_{t}\Psi)^2+(\partial_{\theta}\Psi)^2]
\nonumber\\
&&{}+\frac{1}{2}f^{-1}f^{IJ}
\left[\left(\partial_{t}Q_{I}+\Psi\partial_{t}\Phi_{I}-\Phi_I\partial_{t}\Psi\right)
\left(\partial_{t}Q_{J}+\Psi\partial_{t}\Phi_{J}-\Phi_J\partial_{t}\Psi\right)\right.
\nonumber\\
&&{}\left.+
\left(\partial_{\theta}Q_{I}+\Psi\partial_{\theta}\Phi_{I}-\Phi_I\partial_{\theta}\Psi\right)
\left(\partial_{\theta}Q_{J}+\Psi\partial_{\theta}\Phi_{J}-\Phi_J\partial_{\theta}\Psi\right)\right]\nonumber \\
&&{}+2[(\partial_{t}\phi)^2+(\partial_{\theta}\phi)^2],
\label{lambda-t}
\\
t^{-1}\partial_{\theta}\lambda&=&
\frac{1}{4}f^{-2}\partial_{t}f\partial_{\theta}f
+\frac{1}{4}f^{IJ}f^{MN}\partial_{t}f_{IM}\partial_{\theta}f_{JN}
\nonumber\\
&&{}+e^{-2a\phi}
f^{IJ}\partial_{t}\Phi_{I}\partial_{\theta}\Phi_{J}
+e^{2a\phi}f^{-1}\partial_{t}\Psi\partial_{\theta}\Psi
\nonumber\\
&&{}+\frac{1}{2}f^{-1}f^{IJ}
\left(\partial_{t}Q_{I}+\Psi\partial_{t}\Phi_{I}-\Phi_I\partial_{t}\Psi\right)
\left(\partial_{\theta}Q_{J}+\Psi\partial_{\theta}\Phi_{J}-\Phi_J\partial_{\theta}\Psi\right)\nonumber \\
&&{}+2\partial_{t}\phi\partial_{\theta}\phi.
\label{lambda-theta}
\end{eqnarray}
The integrability condition $\partial_{t}\partial_{\theta}\lambda=\partial_{\theta}\partial_{t}\lambda$ 
of equations (\ref{lambda-t}) and (\ref{lambda-theta})
is assured whenever the other evolution equations are satisfied. 
Note that we can obtain a evolution equation for $\lambda$ from equations for others. 
\begin{eqnarray}
\partial^2_{t}\lambda -\partial^2_{\theta}\lambda
&=&\frac{1}{4}f^{-2}[-(\partial_{t}f)^2+(\partial_{\theta}f)^2]
+\frac{1}{4}f^{IJ}f^{MN}(-\partial_{t}f_{IM}\partial_{t}f_{JN}+\partial_{\theta}f_{IM}\partial_{\theta}f_{JN})\nonumber \\
&&{}+e^{-2a\phi}f^{IJ}(-\partial_{t}\Phi_{I}\partial_{t}\Phi_{J}+\partial_{\theta}\Phi_{I}\partial_{\theta}\Phi_{J})
+e^{2a\phi}f^{-1}[-(\partial_{t}\Psi)^2+(\partial_{\theta}\Psi)^2]
\nonumber\\
&&{}+\frac{1}{2}f^{-1}f^{IJ}
\left[-\left(\partial_{t}Q_{I}+\Psi\partial_{t}\Phi_{I}-\Phi_I\partial_{t}\Psi\right)
\left(\partial_{t}Q_{J}+\Psi\partial_{t}\Phi_{J}-\Phi_J\partial_{t}\Psi\right)\right.
\nonumber\\
&&{}\left.+
\left(\partial_{\theta}Q_{I}+\Psi\partial_{\theta}\Phi_{I}-\Phi_I\partial_{\theta}\Psi\right)
\left(\partial_{\theta}Q_{J}+\Psi\partial_{\theta}\Phi_{J}-\Phi_J\partial_{\theta}\Psi\right)\right]\nonumber \\
&&{}+2[-(\partial_{t}\phi)^2+(\partial_{\theta}\phi)^2].
\label{lambda-evolution-eq}
\end{eqnarray}
Thanks to the areal time coordinate, the metric function $\lambda$ is decoupled with other fields. 
Therefore, it is enough to solve the evolution equations at first. 
After that, we must demand compatibility conditions for $\lambda$. 
That is, equations (\ref{lambda-t}) and (\ref{lambda-theta}) are ones to determine the metric function $\lambda$. 
Under the coordinate (\ref{areal-time-coordinate}), 
we will call a system of equations (\ref{phi-evolution-eq}), (\ref{q-evolution-eq}), (\ref{f-evolution-eq}), 
(\ref{Phi-evolution-eq}), (\ref{Psi-evolution-eq}), (\ref{lambda-t}), (\ref{lambda-theta})
{\it $T^{D-2}$-symmetric EMD system}.


Now, we have the following conclusion by direct calculation: 
\begin{lemma}
\label{wave-map}
Let $M={\rm R}\times T^{2}$ and $N={\rm R}^{\frac{1}{2}(D-2)(D+1)}$ be manifolds 
with Lorentzian metric $\eta$ and Riemannian 
metric $h$, respectively. 
Then, a wave map $U :M\rightarrow N$ is equivalent to the evolution 
equations (\ref{phi-evolution-eq}), (\ref{q-evolution-eq}), (\ref{f-evolution-eq}), 
(\ref{Phi-evolution-eq}), (\ref{Psi-evolution-eq}), 
of the $T^{D-2}$-symmetric EMD system. 
Here, the action for the wave map is 
\begin{eqnarray}
\label{wm-lag}
S_{{\rm WM}}=\int dtd\theta d\psi{\cal L}_{{\rm WM}} 
=\int dtd\theta d\psi\sqrt{-{\rm det}\eta}\eta^{\alpha\beta}h_{AB}\partial_{\alpha}U^{A}\partial_{\beta}U^{B},
\end{eqnarray}
and the metrics are 
\begin{eqnarray}
\eta:=-dt^2+d\theta^2+t^{2}d\psi^2,
\hspace{.5cm}0\leq\theta,\psi\leq2\pi,
\end{eqnarray} 
and 
\begin{eqnarray}
\label{target}
h&:=&\frac{1}{4f^2}df^2+\frac{1}{4}f^{IJ}f^{KL}df_{IK}df_{JL}
+e^{-2a\phi}f^{IJ}d\Phi_{I}d\Phi_{J}+\frac{e^{2a\phi}}{f}d\Psi^2\nonumber \\
&&{}+\frac{1}{2f}f^{IJ}(dQ_{I}+\Psi d\Phi_{I}-\Phi_{I}d\Psi)(dQ_{J}+\Psi d\Phi_{J}-\Phi_{J}d\Psi)
+2d\phi^2.
\end{eqnarray} 
Note that $U$ is independent of $\psi$. 
\hspace*{\fill}$\Box$
\end{lemma}

The integrals (\ref{integral}) are independent of time $t$. 
This fact follows from Noether's theorem~\cite{BCM,SS}. 
Let
$Z$ be a Killing  vector for the metric $h$ given
by (\ref{target}).  
It is well known that the following quantity is independent of the choice of compact Cauchy hypersurfaces,  
\begin{eqnarray}
E(Z,S) = \int_{S} \eta^{\alpha\beta} h_{AB} \frac{\partial U^{A}}{\partial x^{\alpha}}Z^{B}dS_{\beta},
\label{noether}
\end{eqnarray} 
where $\alpha,\beta=t,\theta,\psi$, 
$U^A=(P_{I},Q_{I},\Phi_{I},\Psi,\phi)^T$ and 
$dS_{\alpha} = \partial_{\alpha}\vee (dt\wedge d\theta\wedge d\psi)$, and $\vee$ denotes
contraction. It is easy to see that $\frac{\partial}{\partial Q_{I}}$ are Killing  vectors for the metric $h$.  
Taking $Z= \frac{\partial}{\partial Q_{I}}$ and
$S=\{t={\rm constant}\}=\Sigma$ one obtains the integrals (\ref{integral}). 
Thus, they are conservation quantities when the wave map system holds.

Although Lemma~\ref{wave-map} is quite general and useful, 
geometry of the target space $(N,h)$ is very complicated since scalar multiplet $f_{IJ}$ does not be fixed. 
Therefore, to analyze the system furthermore, 
we assume the following,  
\begin{eqnarray}
\label{diagonal}
f_{IJ}=e^{2P_{I}}\delta_{IJ},
\end{eqnarray} 
and $P=\sum_{I}P_{I}$. 
This assumption does not restrict to diagonal (i.e. polarized) spacetimes. 
Indeed, our $T^{D-2}$-symmetric spacetimes with the assumption 
include four-dimensional {\it unpolarized} Gowdy symmetric ones as a special case. 
Under the assumption (\ref{diagonal}), the metric of the target space can be written by 
\begin{eqnarray}
\label{target-simple}
h&=&dP^2+\sum_{I=3}^{D-1}dP_{I}^2
+e^{-2a\phi}\sum_{I=3}^{D-1}e^{-2P_{I}}d\Phi_{I}^2+e^{-2P+2a\phi}d\Psi^2\nonumber \\
&&{}+\frac{1}{2}e^{-2P}\sum_{I=3}^{D-1}e^{-2P_{I}}(dQ_{I}+\Psi d\Phi_{I}-\Phi_{I}d\Psi)^2
+2d\phi^2.
\end{eqnarray} 
For this target space, we can show the following global existence theorem:
\begin{theorem}
\label{thm1}
Let $({\cal M}_{D}, g_{\mu\nu}, A_{\mu}, \phi)$ be 
the maximal Cauchy development of smooth $T^{D-2}$-symmetric Cauchy data on 
${\cal M}_{D-1}\approx T^{D-1}$ 
for the $T^{D-2}$-symmetric EMD system. 
Then, under the assumption (\ref{diagonal}), 
${\cal M}_{D}$ can be covered by the areal time coordinate with 
$t\in(0,\infty)$.
\end{theorem}

It is known the local existence result for the wave map system 
(e.g. Theorem 7.1 of~\cite{SS}). Therefore, it is enough to show 
boundedness for the sup norm of the zeroth, the first and the second 
derivatives of the functions 
on compact subinterval of $(0,\infty)$ for time coordinate $t$.

Let us define the energy-momentum tensor ${\cal T}_{\alpha\beta}$ associated 
with the 
Lagrangian density (\ref{wm-lag}), which has the form 
\begin{eqnarray}
{\cal T}_{\alpha\beta}:=h_{AB}(\partial_{\alpha}U^{A}
\partial_{\beta}U^{B}
-\frac{1}{2}\eta_{\alpha\beta}\partial_{\alpha}U^{A}
\partial^{\alpha}U^{B}).
\end{eqnarray}
Components of ${\cal T}_{\alpha\beta}$ are as follows, 
with $\langle ,\rangle$ and $\Vert\bullet\Vert$ denoting the inner product 
and the norm with respect to $h$, 

\begin{eqnarray}
{\cal T}_{\alpha\beta}&=&
\left(\begin{array}{ccc}
{\cal T}_{tt} & {\cal T}_{t\theta} & {\cal T}_{t\psi} \\
{\cal T}_{\theta t} & {\cal T}_{\theta\theta} & {\cal T}_{\theta\psi} \\
{\cal T}_{\psi t} & {\cal T}_{\psi\theta} & {\cal T}_{\psi\psi} \\
\end{array}\right)
=\frac{1}{2}
\left(\begin{array}{ccc}
\Vert\partial_{t}U\Vert^{2}+\Vert\partial_{\theta}U\Vert^{2} & 2\langle\partial_{t}U,\partial_{\theta}U\rangle & 0 \\
2\langle\partial_{t}U,\partial_{\theta}U\rangle & 
\Vert\partial_{t}U\Vert^{2}+\Vert\partial_{\theta}U\Vert^{2} & 0 \\
0 & 0 & t^{2}
(\Vert\partial_{t}U\Vert^{2}-\Vert\partial_{\theta}U\Vert^{2}) \\
\end{array}\right)
\nonumber \\
&=:&
\left(\begin{array}{ccc}
{\cal E} & {\cal F} & 0 \\
{\cal F} & {\cal E} & 0 \\
0 & 0 & -t^2{\cal G} \\
\end{array}\right).
\end{eqnarray}
Note that equations for $\lambda$ can be simplify by these quantities as follows:
\begin{eqnarray}
\label{lambda-t'}
\partial_{t}\lambda 
&=&t{\cal E},
\\
\label{lambda-theta'}
\partial_{\theta}\lambda
&=&t{\cal F},
\\
\label{lambda-evolution-eq'}
\partial^2_{t}\lambda -\partial^2_{\theta}\lambda
&=&{\cal G}.
\end{eqnarray}

By using {\it light cone estimate}~\cite{MV} and 
{\it Christodoulou-Tahvildar-Zadeh's identity}~\cite{CTZ}, 
we have the following lemma:
\begin{lemma}[Lemma 2 of~\cite{NM03}]
\label{energy-estimate}
There is a positive constant $C$ such that
\begin{equation}
\label{lemma2}
   {\cal E} \leq C \left[1 + \frac{1}{t^{2}}\right], \quad t \in (0, \infty), \nonumber
\end{equation}
where $C$ depends only on the initial data at $t = t_0$.
\hfill$\Box$
\end{lemma}
\noindent
{\it Proof of Theorem \ref{thm1}}:
From lemma \ref{energy-estimate}, we have the desired bounds on 
$\mid\partial_{t}P\mid$, $\mid\partial_{\theta}P\mid$, 
$\mid\partial_{t}P_{I}\mid$, $\mid\partial_{\theta}P_{I}\mid$, 
$\mid e^{-2(a\phi+P_{I})}\partial_{t}\Phi_{I}\mid$, 
$\mid e^{-2(a\phi+P_{I})}\partial_{\theta}\Phi_{I}\mid$, 
$\mid\partial_{t}\phi\mid$, $\mid\partial_{\theta}\phi\mid$, 
$\mid e^{-2(P-a\phi)}\partial_{t}\Psi\mid$, 
$\mid e^{-2(P-a\phi)}\partial_{\theta}\Psi\mid$, 
$\mid e^{-2(P+P_{I})}(\partial_{t}Q_{I}+\Psi\partial_{t}\Phi_{I}-\Phi_{I}\partial_{t}\Psi)\mid$, 
$\mid e^{-2(P+P_{I})}(\partial_{\theta}Q_{I}+\Psi\partial_{\theta}\Phi_{I}-\Phi_{I}\partial_{\theta}\Psi)\mid$, 
for all $t\in (0,\infty)$.
Once we have bounds on the first derivatives of $P$, $P_{I}$ and $\phi$, 
it follows that $P$, $P_{I}$ and $\phi$ are bound for all $t\in (0,\infty)$. 
Then, we have bounds on $\partial_{t}\Phi_{I}$, $\partial_{\theta}\Phi_{I}$, 
$\partial_{t}\Psi$, $\partial_{\theta}\Psi$, $\partial_{t}Q_{I}+\Psi\partial_{t}\Phi_{I}-\Phi_{I}\partial_{t}\Psi$ and 
$\partial_{\theta}Q_{I}+\Psi\partial_{\theta}\Phi_{I}-\Phi_{I}\partial_{\theta}\Psi$. 
Consequently, $\Phi_{I}$, $\Psi$, $\partial_{t}Q_{I}$ and $\partial_{\theta}Q_{I}$ 
are bounded. 
Finally, we have boundedness on $Q_{I}$.

Next, we must show bounds on the second and higher derivatives of the functions.  
There is a well-known general fact that, in order to ensure the continuation 
of a solution of a system of semi-linear wave equations, 
it is enough to bound the first derivative pointwise.  
Then, we have boundedness of the higher derivatives.

By the constraint equations~(\ref{lambda-t'}) 
and (\ref{lambda-theta'}), 
boundedness and compatibility with the periodicity in $\theta$ for the function $\lambda$ is 
also shown. 
The arguments for that are the same with one of the proof 
of Theorem 1 of \cite{NM03}. 
Thus, we have completed the proof of Theorem~\ref{thm1}.
\hfill$\Box$

\section{Global existence theorem in constant mean curvature (CMC) time coordinate}
To show a global existence theorem in CMC time coordinate for $T^{D-2}$-symmetric spacetimes, 
Henkel's observation is applicable~\cite{HO}. 
He has considered prescribed mean curvature foliations in locally $U(1)\times U(1)$-symmetric 
four-dimensional spacetimes with matter fields.  
The system of our $D$-dimensional spacetimes is 
similar with one treated in that paper\footnote{Rather, our system is simpler than Henkel's 
because there is no shift vector in ours.}. 
One difference between Henkel's and ours is target spaces of the wave maps. 
Fortunately, arguments in the proofs of Propositions 5.4-5.6 of \cite{HO} do not depend on properties of the target spaces. 
In addition, we need not estimate functions of matter fields separately since 
the field equations of the matter fields are included in the wave map system in our case. 
Thus, from negativity of mean curvature of areal time slices, 
\begin{eqnarray}
K=-e^{P-\lambda}\left[\partial_{t}\lambda-\partial_{t}P+\frac{1}{t}\right]\leq -\frac{3e^{P-\lambda}}{4t}<0,
\hspace{.5cm}t\in(0,\infty),
\end{eqnarray} 
(where equation (\ref{lambda-t'}) has used) 
the spacetimes has crushing singularity into the past and thus 
a neighborhood near the singularity can be foliated by compact 
CMC hypersurfaces~\cite{GC}. 
Once we have one compact CMC hypersurface with negative mean curvature, 
it is shown that the CMC foliation covers the entire future of the initial 
CMC hypersurface (see \cite{ARR}). 
Thus, the following theorem is obtained:
\begin{theorem}
\label{cmc}
Let $({\cal M}_{D}, g_{\mu\nu}, A_{\mu}, \phi)$ be 
the maximal Cauchy development of smooth $T^{D-2}$-symmetric Cauchy data on ${\cal M}_{D-1}\approx T^{D-1}$ 
for the $T^{D-2}$-symmetric EMD system. 
Then, under the assumption (\ref{diagonal}), 
${\cal M}_{D}$ can be covered by hypersurfaces of the constant mean curvature in the range $(-\infty,0)$.
\hfill$\Box$
\end{theorem}

\section{Existence of asymptotically velocity-terms dominated (AVTD) solutions}
\subsection{Analytic case}
\subsubsection{Generalities}
As a method to construct AVTD singular solutions, the 
{\it Fuchsian algorithm} has been 
developed~\cite{KR}. 
To answer the question rigorously whether the AVTD behavior exists or not 
for vacuum Einstein gravity in arbitrary spacetime dimension, 
the algorithm will be used.

Now, we briefly review the Fuchsian algorithm.  
Let us consider a hyperbolic PDE system,  
\begin{equation}
\label{f=0}
F\left[u(t,x^{\alpha})\right]=0.
\end{equation}
Generically, $u$ can have any number of components.  Here, we will assume 
that the PDE is singular with respect to the argument $t$.  
The Fuchsian algorithm consists of three steps:  
At first, identify the leading (singular) 
terms $u_{0}(t,x)$ which are parts of the desired 
expansion for $u$.  This means that the most singular terms cancel each other 
when $u_{0}(t,x)$ is substituted in equation~(\ref{f=0}).  
One can get the solution $u_{0}$ from a system of 
velocity-terms-dominated (VTD) equations which are 
obtained by neglecting spatial derivative terms from 
full equations (\ref{f=0}). 
Second, introduce a renormalized unknown function $v(t,x)$, which is given by 
\begin{equation}
\label{u=a+tv}
u=u_{0}+t^{m}\tilde{u}.
\end{equation}
If $u_{0}\sim t^{k}$, we should set $m=k+\varepsilon$, 
where $\varepsilon >0$.  
Thus, $\tilde{u}$ is a regular part of the desired 
expansion for $u$.  
Finally, obtain a {\it Fuchsian system} for $\tilde{u}$ by substituting 
equation~(\ref{u=a+tv}) 
in equation~(\ref{f=0}).  That is, 
\begin{equation}
\label{fuchsian-system}
({\cal D}+N(x))\tilde{u}=t^{\alpha}V(t,x,\tilde{u},\partial_{x}\tilde{u}), 
\end{equation}
where ${\cal D}:= t\partial _{t}$ and $N$ is a matrix which is independent of 
$t$ and $\alpha >0$.  
Note that one can always take $\alpha =1$ 
by introducing $t^{\alpha}$ as a new time variable. 
$V$ can be assumed to be analytic in all of arguments except $t$ 
and continuous in $t$ since 
the following existence theorem (Theorem~\ref{fuchsian-theorem}) 
is a {\it singular version of the Cauchy-Kowalewskaya theorem} essentially. 
Note that equation~(\ref{fuchsian-system}) is a {\it singular} 
PDE system for the 
{\it regular} function $v$.  

Once we have the Fuchsian system, we can show the existence of a unique 
solution for prescribed singular part $u_{0}$ by the following theorem.  
\begin{theorem}[Theorem 3 of~\cite{KR}]
\label{fuchsian-theorem}
Let us consider a system (\ref{fuchsian-system}), 
where $N$ is an analytic matrix near $x=0$, such that $\|\sigma^{N}\|\leq C$ 
for $0<\sigma <1$ {\rm (boundedness condition)} 
and $V$ is analytic in space $x$ and continuous in time $t$.  
Then the Fuchsian system (\ref{fuchsian-system}) has a unique solution 
which is 
defined near $x=0$ and $t=0$, and which is analytic in space $x$ 
and continuous 
in time $t$, and tend to zero as $t\rightarrow 0$.  
\hfill$\Box$
\end{theorem}
Note that the boundedness condition holds if every eigenvalue of 
$A$ is non-negative.  
Theorem~\ref{fuchsian-theorem} implies that 
renormalized unknown functions must vanish as $t\rightarrow 0$ 
if the conditions are satisfied.  
Therefore, the only singular terms which are solutions to VTD 
equations remain and they are solutions to full field equations at $t=0$.  
Thus, we obtain AVTD singular solutions.  

\subsubsection{Application to vacuum $T^{D-2}$-symmetric spacetimes}
As mentioned in Section~\ref{intro}, it is interesting for us in the vacuum 
case. 
Let us take $\Phi_{I}\equiv \Psi\equiv \phi\equiv a\equiv 0$. 
Then, we obtain 
a metric of the target space of the wave map from (\ref{target-simple}), 
\begin{eqnarray}
\label{target-simple-vacuum}
h=dP^2+\sum_{I=3}^{D-1}dP_{I}^2+\frac{1}{2}e^{-2P}\sum_{I=3}^{D-1}e^{-2P_{I}}dQ_{I}^2.
\end{eqnarray} 
By Lemma~\ref{wave-map} (or equations (\ref{f-evolution-eq}) 
and (\ref{q-evolution-eq})\footnote{In this case, the dilaton and the Maxwell equations 
(\ref{phi-evolution-eq}), (\ref{Phi-evolution-eq}), (\ref{Psi-evolution-eq}) are 
automatically satisfied.}) 
we have a system of evolution equations for vacuum Einstein gravity as follows:
\begin{eqnarray}
\label{gowdy-eq-p}
{\cal D}^2P_{I}-t^2\partial_{\theta}^2P_{I}
&=&-\frac{1}{2}e^{-2(P+P_{I})}[({\cal D}Q_{I})^2-(t\partial_{\theta}Q_{I})^2], 
\\
\label{gowdy-eq-q}
{\cal D}^2Q_{I}-t^2\partial_{\theta}^2Q_{I}
&=&2[{\cal D}(P+P_{I}){\cal D}Q_{I}-t^2(\partial_{\theta}P+\partial_{\theta}P_{I})\partial_{\theta}Q_{I}].
\end{eqnarray} 
Neglecting spatial derivative terms from equations (\ref{gowdy-eq-p})-(\ref{gowdy-eq-q}), one can obtain VTD equations as follows:
\begin{eqnarray}
\label{vtd-eq-p}
{\cal D}^2P_{I}
&=&-\frac{1}{2}e^{-2(P+P_{I})}({\cal D}Q_{I})^2,
\\
\label{vtd-eq-q}
{\cal D}^2Q_{I}
&=&2{\cal D}(P+P_{I}){\cal D}Q_{I}.
\end{eqnarray} 
From these equations (\ref{vtd-eq-p})-(\ref{vtd-eq-q})
we can find VTD solutions as follows:  
\begin{eqnarray}
\label{vtd-sol-p}
P^{{\rm VTD}}_{I}(t,\theta)&=&p_{I0}(\theta)\ln t+p_{I1}(\theta),\\
\label{vtd-sol-q}
Q^{{\rm VTD}}_{I}(t,\theta)&=&q_{I0}(\theta)+t^{2(p_{0}+p_{I0})}q_{I1}(\theta),
\end{eqnarray} 
where $p_{0}:=\sum_{I}p_{I0}$. Hereafter we put $k_{I}:=p_{0}+p_{I0}$. 
Thus, the following formal solutions which have the leading terms 
$P^{{\rm VTD}}_{I}$ and $Q^{{\rm VTD}}_{I}$ are obtained: 
\begin{eqnarray}
\label{formal-sol-p}
P_{I}(t,\theta)&=&P^{{\rm VTD}}_{I}+t^{\epsilon_{I}}\pi_{I}(t,\theta),\\
\label{formal-sol-q}
Q_{I}(t,\theta)&=&Q^{{\rm VTD}}_{I}+t^{2k_{I}}\kappa_{I}(t,\theta),
\end{eqnarray} 
where $\epsilon_{I} >0$ and $k_{I}>0$.  

Let us be 
$\bar{\pi}_{I}:=(\pi_{0I},\pi_{1I},\pi_{2I})
:=(\pi_{I},{\cal D}\pi_{I},t\partial_{\theta}\pi_{I})$ 
and $\bar{\kappa}:=(\kappa_{0I},\kappa_{1I},\kappa_{2I})
:=(\kappa_{I},{\cal D}\kappa_{I},t\partial_{\theta}\kappa_{I})$.   
From these, we define 
$v:=(\bar{\pi}_{I},\bar{\kappa}_{I})^{T}$. 
We have a Fuchsian system for $v$ 
by inserting solutions (\ref{formal-sol-p}) and (\ref{formal-sol-q}) 
into the system (\ref{gowdy-eq-p}) and (\ref{gowdy-eq-q}): 
\begin{eqnarray}
\label{fuchsian-eq}
({\cal D}+N)v=t^{\delta}V(t,\theta,v,\partial_{\theta}v), 
\end{eqnarray} 
where $\delta 
=\min\{2k_{I}-\epsilon_{I},2(1-k_{I})-\epsilon_{I},\epsilon_{I}\}$ 
for any $I$, 
$V$ is a vector-valued regular function and 
$N$ is 
\begin{eqnarray}
N=
\left[\begin{array}{cccccc}
N_{\epsilon_{3}} &        &                    &           &        &             \\
                 & \ddots &                    &           &        &             \\
                 &        & N_{\epsilon_{D-1}} &           &        &             \\
                 &        &                    & N_{k_{3}} &        &             \\
                 &        &                    &           & \ddots &             \\
                 &        &                    &           &        & N_{k_{D-1}} \\ 
\end{array}\right],
\end{eqnarray}
where 
\begin{eqnarray}
N_{\epsilon_{I}}:=
\left[\begin{array}{ccc}
0 & -1 & 0 \\
\epsilon_{I}^2 & 2\epsilon_{I} & 0 \\
0 & 0 & 0 \\
\end{array}\right]
\hspace{.5cm} {\rm and} \hspace{.5cm}
N_{k_{I}}:=
\left[\begin{array}{ccc}
0 & -1 & 0 \\
0 & 2k_{I} & 0 \\
0 & 0 & 0 \\
\end{array}\right],
\end{eqnarray}
which are independent of $t$. It is easy to see that 
every eigenvalue of $N$ is non-negative and regularity of 
the right-hand-side of 
the system (\ref{fuchsian-eq}) holds if and only if 
$0<\epsilon_{I}<\min\{2k_{I}, 2(1-k_{I})\}$, i.e. $0<k_{I}<1$.  
Thus, we have the following theorem from Theorem~\ref{fuchsian-theorem}. 
\begin{theorem}
\label{avtd-th-analytic}
Let $p_{IA}(\theta)$, $q_{IA}(\theta)$, where $A=0$, $1$, be real 
analytic functions of $\theta$ in a neighborhood of $\theta =\theta_{0}$ and 
$0<k_{I}<1$ for all $\theta$. 
Then, for the vacuum Einstein equations (\ref{gowdy-eq-p}) 
and (\ref{gowdy-eq-q}) 
in arbitrary dimensions ($D\geq 4$), 
there exists a unique solution with the form 
(\ref{formal-sol-p}) and (\ref{formal-sol-q}) 
in a neighborhood of $\theta =\theta_{0}$, where 
$\pi_{I}$, $\kappa_{I}$ tend to zero as 
$t\rightarrow 0$.  
\hfill$\Box$
\end{theorem}
Theorem~\ref{avtd-th-analytic} implies that VTD solutions 
$P^{{\rm VTD}}_{I}(t,\theta)$ 
and $Q^{{\rm VTD}}_{I}(t,\theta)$ 
are solutions to full Einstein equations (\ref{gowdy-eq-p}) 
and (\ref{gowdy-eq-q}) near the initial singularity. 
The VTD solutions depend on the maximal number (i.e. $4\times(D-3)$)  
of arbitrary functions $p_{IA}(\theta)$ and $q_{IA}(\theta)$. 
Thus, it has been shown that $T^{D-2}$-symmetric vacuum spacetimes 
on $T^{D-1}\times{\rm R}$ have 
AVTD behavior in general. 

\subsection{Smooth case}
\subsubsection{Generalities}
The argument in this subsection has been developed by Rendall~\cite{RA} 
(which was generalized in~\cite{SF}). We will review it, which is a 
{\it smooth version of the Fuchsian algorithm}. 

We want to extend to the $C^{\infty}$ category of the existence of AVTD 
singular solutions. To do this, we introduce notions of regularity and 
formal solutions of equation (\ref{fuchsian-system}). 
\begin{definition}
\label{def-regularity}
A function $f(t,x)$ from an open subset
$\Omega\subset[0,\infty)\times{\rm R}^n$ to ${\rm R}^m$ is called
\emph{regular} if it is $C^\infty$ for all $t>0$ and if its partial
derivatives of any order with respect to $x\in{\rm R}^n$ extend
continuously to $t=0$ (within $\bar{\Omega}$).
\end{definition}

\begin{definition}
\label{def-formal-sol}
A finite sequence $(v_1,\dots,v_p)$ of functions defined on an
open subset $\Omega\subset[0,\infty)\times{\rm R}^n$ is called a
\emph{formal solution of order $p$} of the differential equation
(\ref{fuchsian-system}) on $\Omega$ if 
\begin{enumerate}
\item each $v_i$ is regular and
\item $({\cal D}+N(x))v_i - tV(t,x,v_i,\partial_{x}v_{i})
         = O(t^i)$ for all $i$ as $t\to0$ in $\Omega$, where $O$-symbol is 
taken in the sense of uniform convergence on  compact subsets. 
\end{enumerate}
\end{definition}

If functions $N$ and $V$ in (\ref{fuchsian-system}) are analytic ones, 
existence of analytic solutions can be shown~\cite{KR}. 
Contrary, if $N$ and $V$ are merely smooth, the following existence theorem 
for formal solutions of any order is obtained:
\begin{lemma}[Lemma 2.1 of ~\cite{RA}]
\label{lemma-formal-sol}
If $V$ is regular and $N$ is smooth and satisfies
$\Vert\sigma^{N}\Vert\le{C}$ for some constant $C$ and for $0<\sigma <1$ 
in a neighborhood of zero, 
then equation (\ref{fuchsian-system}) has a
formal solution of any given order which vanishes at $t=0$.
\hfill$\Box$
\end{lemma}
The Fuchsian system (\ref{fuchsian-eq}) 
satisfies the conditions in Lemma \ref{lemma-formal-sol} if 
$\epsilon_{I}<2k_{I}$ and $\epsilon_{I}<2-2k_{I}$. 
Then, this system has a formal solution of any order which vanishes at 
$t=0$.

Now, we will rewrite the system 
(\ref{fuchsian-system}) to a symmetric hyperbolic system defined as below: 
\begin{definition}
\label{def-reg-sh}
We say that the system of differential equations
\begin{equation}
\label{shf}
\{{\cal A}^0(t,x){\cal D}+{\cal N}(x)+t{\cal A}^j(t,x,v)\partial_{j}\}v 
= t {\cal V}(t,x,v)
\end{equation}
is \emph{regular symmetric hyperbolic (RSH)} if ${\cal A}^0$ is uniformly
positive definite and symmetric, the ${\cal A}^j$ are symmetric, and all
coefficients are assumed to be regular.
\end{definition}
For generic symmetric hyperbolic systems, it is a key point that 
energy inequalities defined in the $L^2$-Sobolev space hold. 
The basic idea used to prove local existence results for 
symmetric hyperbolic systems is as follows. 
A family of approximated systems is constructed and is solved, and then, 
$L^2$ energy estimates are used to show that solutions of the approximated 
systems converge to a solution of the original systems 
in a certain limit~\cite{FR,MS,TM}. According to this, 
a sequence of analytic data as smooth and approximate data was 
used and convergence of the corresponding analytic solutions to 
a smooth solution was shown for the RSH system (\ref{shf}) coming from 
the Fuchsian system (\ref{fuchsian-system}) in~\cite{RA}. 

It is impossible to get a tractable 
RSH system from the Fuchsian system in any cases, 
since when the original Fuchsian system would be rewritten 
to a symmetric hyperbolic system, 
two disadvantages may happen. One is violation of the boundedness condition 
for ${\cal N}$ (positivity of eigenvalues of ${\cal N}$) 
and another is change of powers of $t$ in the 
right-hand-side of the system. To overcome the first problem, 
we will use the notion of formal solutions (and Lemma \ref{lemma-formal-sol}) 
and, for the second problem, we will consider as the cases may be.

A formal solution of order $p$ of the RSH system 
is defined as Definition~\ref{def-formal-sol}. 
Note that it is necessary to verify that a formal solution of 
(\ref{fuchsian-eq}) of order $p$ is also a formal solution of 
(\ref{shf}) of order $p$. If this can be done, 
we can proceed the next step.

Given a formal solution $\{ v_{1},\dots ,v_{i}\}$ of (\ref{fuchsian-system}), 
it is possible to consider the difference between a genuine solution $v$ 
(which has been known yet) and the formal solution. 
Put $z_{i}:=t^{1-i}(v-v_{i})$. Then, 
we can obtain a system for $z_{i}$ as follows: 
\begin{eqnarray}
\label{fuchsian-eq-3rd}
\{{\cal A}^0(t,x){\cal D}+[{\cal N}(x)+(i-1){\cal A}^0(t,x)]
+ t {\cal A}^j(t,x,w_{i}+t^{i-1}z_{i})\partial_{j}\}z_{i} 
= t {\cal V}_{i}(t,x,z_{i}),
\end{eqnarray} 
for some regular function ${\cal V}_{i}$. 
Note that choosing $i$ large enough makes real part of the eigenvalues of 
${\cal N}+(i-1){\cal A}^0$ positive, since ${\cal A}^0$ 
is uniformly positive definite. 

If the data are approximated by a sequence of analytic data 
$S_{m}:=(p_{IAm}, q_{IAm})$, 
a corresponding sequence of analytic solution is obtained. 
We can construct the sequence $S_{m}$ which converges to smooth data 
$S:=(p_{IA}, q_{IA})$ in $C^{\infty}$, uniformly on compact subsets. 
If the (approximate) formal solutions are constructed as in the proof 
of Lemma \ref{lemma-formal-sol}, then $v_{mi}\rightarrow v_{i}$ as 
$m\rightarrow\infty$, uniformly on compact subsets. Here $v_{mi}$ are 
analytic formal solutions of order $i$ corresponding to the 
analytic data $S_{m}$ and $v_{i}$ is a formal solution order $i$ 
for the smooth data $S$. 
The same is true for the spatial derivatives of these functions of any order. 
Therefore, one can conclude that the sequence of coefficients of the system 
(\ref{fuchsian-eq-3rd}) also converge on compact subsets as 
$m\rightarrow\infty$. 

Now, our task will be show that the sequence of solutions $v_{mi}$ exists 
for a time interval independent of $m$ and has a limit as $m\rightarrow\infty$ 
solving the system (\ref{shf}). 
Fortunately, 
the global existence theorem (Theorem \ref{thm1}) implies that there is 
a sequence of smooth solutions to (\ref{fuchsian-eq-3rd}) 
on a common time interval for all $m$. 
Then, the following theorem ensures existence of genuine smooth solutions. 
\begin{theorem}[Section 4 of~\cite{RA} and Theorem 5.1 of~\cite{SF} 
(see also Theorem 2.3 of~\cite{KS})]
\label{fuchsian-theorem-smooth}
Let $z_m(t,x)$ be a sequence of regular solutions on
$[0,t_1)\times{\cal U}\subset[0,\infty)\times{\rm R}^n$, with
$z_m(0,x)=0$, to a sequence of RSH equations
\begin{equation}
\label{rshf}
\{{\cal A}_m^0(t,x){\cal D}+ {\bf N}_m(t,x)
+ t {\cal A}_m^j(t,x,z_m)\partial_{j}\}z_{m}
= t {\cal V}_m(t,x,z_m).
\end{equation}
Suppose that ${\bf N}_m$ is positive definite for each $m$ and that the 
coefficients converge uniformly to ${\cal A}_{\infty}^0$, ${\bf N}_{\infty}$, 
${\cal A}_{\infty}^j$ and ${\cal V}_{\infty}$
on compact subsets as $m\rightarrow\infty$, with the same properties as 
${\cal A}_m$,
${\bf N}_m$, ${\cal A}_m^j$ and ${\cal V}_m$, 
and that the corresponding spatial
derivatives converge uniformly as well. Then $z_m$ converges to a
regular solution $z_{\infty}$ of the corresponding system with coefficients
${\cal A}_{\infty}^0$, ${\bf N}_{\infty}$, ${\cal A}_{\infty}^j$ 
and ${\cal V}_{\infty}$ on $[0,t_0)\times{\cal U}$ for some
$t_0$, and $z_{\infty}(0,x)=0$.
\hfill$\Box$
\end{theorem}
Here, we have fixed a value of $i$ large enough and omitted the index $i$. 

Finally, we must verify that the obtained smooth solution 
of (\ref{shf}) is also a solution 
to (\ref{fuchsian-eq}). 
When the above arguments be done, 
the proof of the existence of smooth VTD solutions 
completes.

\subsubsection{Low velocity case}
\label{low-velocity-case}
Now, we will rewrite the Fuchsian system 
(\ref{fuchsian-eq}) to a RSH system as follows: 
\begin{eqnarray}
\label{sh-fuchsian-eq}
({\cal D}+{\cal N}+t{\cal A}^{\theta}\partial_{\theta})v=
t^{\zeta}{\cal V}(t,\theta,v,\partial_{\theta}v), 
\end{eqnarray} 
where $\zeta =\min\{2k_{I}-\epsilon_{J},2(1-k_{I})-\epsilon_{J}, 
1-2k_{I}+\epsilon_{J},\epsilon_{J}\}$ 
for any $I$ and $J$, ${\cal V}$ is a vector-valued regular functions 
and  
\begin{eqnarray}
{\cal N}=
\left[\begin{array}{cccccc}
{\cal N}_{\epsilon_{3}} &        &                    &           &        &             \\
                 & \ddots &                    &           &        &             \\
                 &        & {\cal N}_{\epsilon_{D-1}} &           &        &             \\
                 &        &                    & {\cal N}_{k_{3}} &        &             \\
                 &        &                    &           & \ddots &             \\
                 &        &                    &           &        & {\cal N}_{k_{D-1}} \\ 
\end{array}\right],
\end{eqnarray}
where
\begin{eqnarray}
\label{n}
{\cal N}_{\epsilon_{I}}:=
\left[\begin{array}{ccc}
0 & -1 & 0 \\
\epsilon_{I}^2 & 2\epsilon_{I} & 0 \\
0 & 0 & -1 \\
\end{array}\right]
\hspace{.5cm} 
{\rm and} \hspace{.5cm}
{\cal N}_{k_{I}}:=
\left[\begin{array}{ccc}
0 & -1 & 0 \\
0 & 2k_{I} & 0 \\
0 & 0 & -1 \\
\end{array}\right],
\end{eqnarray}
and
\begin{eqnarray}
{\cal A}^{\theta}=
\left[\begin{array}{ccc}
{\cal A}_{{\rm Her}} &        &   \\
  & \ddots &   \\
  &        & {\cal A}_{{\rm Her}} \\
\end{array}\right],
\hspace{.5cm} {\rm where} \hspace{.5cm}
{\cal A}_{{\rm Her}}:=
\left[\begin{array}{ccc}
0 & 0 & 0 \\
0 & 0 & -1 \\
0 & -1 & 0 \\
\end{array}\right],
\end{eqnarray}
which are independent of $t$. In our case, ${\cal A}^0$ becomes the identity. 
The regularity of the right-hand-side of (\ref{sh-fuchsian-eq}) 
is satisfied if $0<k_{I}<3/4$ for any $I$.

We must verify that solutions to the Fuchsian system (\ref{fuchsian-eq}) 
are also solutions to the RSH system 
(\ref{sh-fuchsian-eq}). 
The system (\ref{sh-fuchsian-eq}) is obtained by replacing 
$t\partial_{\theta}\pi_{0I}$ and $t\partial_{\theta}\kappa_{0I}$ with 
$\pi_{2I}$ and $\kappa_{2I}$, respectively. 
For analytic solutions, the system (\ref{fuchsian-eq}) implies 
${\cal D}(t\partial_{\theta}\pi_{0I}-\pi_{2I})=0$ and 
${\cal D}(t\partial_{\theta}\kappa_{0I}-\kappa_{2I})=0$. 
Since $t\partial_{\theta}\pi_{0I}-\pi_{2I}=0$ and 
$t\partial_{\theta}\kappa_{0I}-\kappa_{2I}=0$ at $t=0$, 
we can conclude that these equations are satisfied in all time. 
Thus, an analytic solution of (\ref{fuchsian-eq}) is an analytic solution of 
(\ref{sh-fuchsian-eq}). 
Repeating this argument for formal solutions of order $i$ 
to the Fuchsian system (\ref{fuchsian-eq}), we have the same conclusion for 
formal solutions. 

Next, we want to apply Theorem \ref{fuchsian-theorem-smooth} 
to our system (\ref{sh-fuchsian-eq}). 
As we can see from (\ref{fuchsian-eq-3rd}) and (\ref{n}), 
it is enough to take $i\geq 3$, where $i$ is the order for the 
formal solution.  
Then, we have a smooth solution to the RSH system (\ref{sh-fuchsian-eq}). 

Finally, we have to show that the obtained solution to (\ref{sh-fuchsian-eq}) 
is also a solution to the Fuchsian system (\ref{fuchsian-eq}). 
From (\ref{sh-fuchsian-eq}) and the fact that $\bar{\pi}$ and 
$\bar{\kappa}$ are vanish at $t=0$, we have 
$\pi_{2I}-t\partial_{\theta}\pi_{0I}=c^{\pi}_{I}(\theta)t$ and 
$\kappa_{2I}-t\partial_{\theta}\kappa_{0I}=c^{\kappa}_{I}(\theta)t$ 
for some functions $c^{\pi}_{I}(\theta)$ and $c^{\kappa}_{I}(\theta)$. 
Suppose analytic functions converging to $\bar{\pi}$ and 
$\bar{\kappa}$ be $\bar{\pi}_{m}$ and $\bar{\kappa}_{m}$. 
Then, we have 
\begin{eqnarray}
\sup_{\theta}\mid c^{\pi}_{I}(\theta)t\mid
&=&\sup_{\theta}\mid(\pi_{2I}-t\partial_{\theta}\pi_{0I})
-(\pi_{2Im}-t\partial_{\theta}\pi_{0Im})\mid\nonumber \\
&\leq&\sup_{\theta}\mid\pi_{2I}-\pi_{2Im}\mid+
\sup_{\theta}t\mid\partial_{\theta}\pi_{0I}
-\partial_{\theta}\pi_{0Im}\mid\rightarrow 0,
\end{eqnarray}
for any $t$ as $m\rightarrow\infty$. Here, equations 
$\pi_{2Im}-t\partial_{\theta}\pi_{0Im}=0$ for any $m$ has been used.  
Thus, $\pi_{2I}-t\partial_{\theta}\pi_{0I}=0$ was shown for all time. 
$\kappa_{2I}-t\partial_{\theta}\kappa_{0I}=0$ can be also proved similarly. 
Thus, it was shown that 
the smooth solution of the RSH system 
(\ref{sh-fuchsian-eq}) is also a smooth 
solution to the original Fuchsian system (\ref{fuchsian-eq}).  To sum up, 
\begin{theorem}
\label{avtd-th-smooth-low}
Suppose that $p_{IA}$ and $q_{IA}$ are smooth functions of
$\theta$ and $0<k_{I}<3/4$ for all $\theta$. 
Then, for each spatial point $\theta =\theta_{0}$, 
there exists a solution of the 
Einstein equations (\ref{gowdy-eq-p}) and (\ref{gowdy-eq-q}) 
in a neighborhood of
$\theta=\theta_{0}$ of the form (\ref{formal-sol-p}) 
and (\ref{formal-sol-q}) where
$2k_{I}-1<\epsilon_{J}<\min\{2k_{I},2-2k_{I}\}$ for any $I$ and $J$, and 
$\pi_{I}$ and $\kappa_{I}$ are regular and tend
to $0$ as $t\rightarrow 0$. Given the form of the expansion and a choice of
$\epsilon_{I}$, the solution is unique.
\hfill$\Box$
\end{theorem}

\subsubsection{Intermediate velocity case}
We want to complement the whole range $0<k_{I}<1$ arising 
in Theorem~\ref{avtd-th-analytic}.  
Now, the following candidates for solutions to the vacuum 
Einstein equations will be considered:  
\begin{eqnarray}
\label{formal-sol-p'}
P_{I}(t,\theta)&=&P^{{\rm VTD}}_{I}+\alpha_{I}(\theta) t^{2-2k_{I}}
+t^{2-2k_{I}+\epsilon_{I}}\pi_{I}(t,\theta),\\
\label{formal-sol-q'}
Q_{I}(t,\theta)&=&Q^{{\rm VTD}}_{I}+t^{2k_{I}}\kappa_{I}(t,\theta),
\end{eqnarray} 
where $\epsilon_{I} >0$ and $k_{I}>0$ and $\alpha_{I}$ are chosen 
as follows:
\begin{eqnarray}
\label{alpha-i}
\alpha_{I}=\frac{1}{2}\left(\frac{\partial_{\theta}q_{I0}}{2-2k_{I}}\right)^2
\exp\left[-2(p_{I1}+\sum_{J=3}^{D-1}p_{J1})\right].
\end{eqnarray} 
Therefore, the number of arbitrary functions is $4\times(D-3)$ still. 
Solutions (\ref{formal-sol-q'}) are the same with (\ref{formal-sol-q}). 
Substituting (\ref{formal-sol-p'}) and (\ref{formal-sol-q'}) into the 
vacuum Einstein equations (\ref{gowdy-eq-p}) and (\ref{gowdy-eq-q}), 
the following RSH system is obtained: 
\begin{eqnarray}
\label{sh-fuchsian-eq'}
({\cal D}+\tilde{{\cal N}}+t{\cal A}^{\theta}\partial_{\theta})v=
t^{\tilde{\zeta}}\tilde{{\cal V}}(t,\theta,v,\partial_{\theta}v), 
\end{eqnarray} 
where $\tilde{\zeta} =\min\{2-2k_{I}-\epsilon_{J},
3-2(k_{I}+k_{J})+\epsilon_{J}, 
2k_{I}-1-\epsilon_{J}\}$ 
for any $I$ and $J$, $\tilde{{\cal V}}$ is a vector-valued regular functions 
and  
\begin{eqnarray}
\tilde{{\cal N}}=
\left[\begin{array}{cccccc}
\tilde{{\cal N}}_{\epsilon_{3}} &        &                    &           &        &             \\
                 & \ddots &                    &           &        &             \\
                 &        & \tilde{{\cal N}}_{\epsilon_{D-1}} &           &        &             \\
                 &        &                    & {\cal N}_{k_{3}} &        &             \\
                 &        &                    &           & \ddots &             \\
                 &        &                    &           &        & {\cal N}_{k_{D-1}} \\ 
\end{array}\right],
\end{eqnarray}
where
\begin{eqnarray}
\tilde{{\cal N}}_{\epsilon_{I}}:=
\left[\begin{array}{ccc}
0 & -1 & 0 \\
(2-2k_{I}+\epsilon_{I})^2 & 2(2-2k_{I}+\epsilon_{I}) & 0 \\
0 & 0 & -1 \\
\end{array}\right],
\end{eqnarray}
which are independent of $t$. 
From $\tilde{\zeta}>0$, we have a regularity condition $1/2<k_{I}<5/6$.
Note that the choice for $\alpha_{I}$ (\ref{alpha-i}) 
is to eliminate the leading order terms in the 
RSH system (\ref{sh-fuchsian-eq'}) obtained after using the solutions 
(\ref{formal-sol-p'}) and (\ref{formal-sol-q'}). 

The procedure of replacement 
($t\partial_{\theta}\pi_{0I}\leftrightarrow\pi_{2I}$ and 
$t\partial_{\theta}\kappa_{0I}\leftrightarrow\kappa_{2I}$) 
is the same with one of the low velocity case 
(Subsection~\ref{low-velocity-case}). 
Thus, we have the following result: 
\begin{theorem}
\label{avtd-th-smooth-int}
Suppose that $p_{IA}$ and $q_{IA}$ are smooth functions of
$\theta$ and $1/2<k_{I}<5/6$ for all $\theta$. 
Then, for each spatial point $\theta =\theta_{0}$, 
there exists a solution of the 
Einstein equations (\ref{gowdy-eq-p}) and (\ref{gowdy-eq-q}) 
in a neighborhood of
$\theta=\theta_{0}$ of the form (\ref{formal-sol-p'}) and 
(\ref{formal-sol-q'}) where
$2(k_{I}+k_{J})-3<\epsilon_{J}<\min\{2k_{I}-1, 2-2k_{I}\}$ 
for any $I$ and $J$, and 
$\pi_{I}$ and $\kappa_{I}$ are regular and tend
to $0$ as $t\rightarrow 0$. Given the form of the expansion and a choice of
$\epsilon_{I}$, the solution is unique.
\hfill$\Box$
\end{theorem}
\begin{remark}
It is impossible to cover the whole range $0<k_{I}<1$ in the analytic case 
(Theorem~\ref{avtd-th-analytic}) by Theorem~\ref{avtd-th-smooth-low} and 
Theorem~\ref{avtd-th-smooth-int}, which are of smooth cases. 
Fortunately, we can overcome this problem by repeating the above method 
$n$ times. The argument is the same with Section 5.7 of \cite{SF}, 
so the details will be omitted. 
If solutions (\ref{formal-sol-p'}) are replaced by
\begin{eqnarray}
\label{formal-sol-p''}
P_{I}(t,\theta)=P^{{\rm VTD}}_{I}+
\sum^{n}_{j=1}\alpha_{Ij}(\theta) t^{(2-2k_{I})j}
+t^{(2-2k_{I})n+\epsilon_{I}}\pi_{I}(t,\theta),
\end{eqnarray}
the regularity condition of the right-hand-side of the symmetric 
hyperbolic system becomes 
\begin{eqnarray}
1-2[(n+1)-(k_{I}+nk_{J})]<\epsilon_{J}<\min\{1-2n(1-k_{I}),2-2k_{I}\},
\end{eqnarray}
for any $I$ and $J$. Here, each $\alpha_{Ij}$ is defined as 
the leading order terms are canceled at each stage like (\ref{alpha-i}). 
Therefore, 
the number of free functions remain $4\times(D-3)$ still. 
Then, we have the following inequality for $k_{I}$:
\begin{eqnarray}
1-\frac{1}{2n}<k_{I}<1-\frac{1}{2(n+2)}.
\end{eqnarray}
Thus, the range $(1/2,1)$ is covered by the infinite sequence of intervals 
$(1-1/2n,1-1/(2(n+2)))$. 
By combining this result with Theorem~\ref{avtd-th-smooth-low},
an existence theorem of smooth AVTD solutions 
under the condition $0<k_{I}<1$ is shown. 
\hfill$\Box$
\end{remark}
\section{Conclusion}
We would like to comment on the structure of the spacetime 
into the future direction since 
we have not discussed on it. 
From point of view of the SCC, 
we want to show future completeness of any causal geodesic of 
the spacetime. 
Recently, it has been shown that 
Gowdy and $U(1)$-symmetric (in the case of small initial data) spacetimes 
are geodesically future complete~\cite{CB,CBC,CBM01,RH}. 
One of key ingredients 
is to show energy decay by using 
{\it corrected energy method}. 
Fortunately, estimate of energy decay for our wave map can be 
shown~\cite{NM04}. 
Another ingredient is the geometric structure of target spaces. 
Unfortunately, our understanding of the structure of (\ref{target}) 
is entirely out of reach at the present time.

\begin{center}
{\bf Acknowledgments}
\end{center}
I am grateful to Lars Andersson, Alan Rendall and 
Yoshio Tsutsumi for commenting on the manuscript.


\end{document}